\documentstyle[twoside,fleqn,espcrc2,epsf]{article}

\newcommand{\la}{\raise.16ex\hbox{$\langle$}}
\newcommand{\ra}{\raise.16ex\hbox{$\rangle$}}
\newcommand{\be}{\begin{equation}}
\newcommand{\ee}{\end{equation}}
\newcommand{\beq}{\begin{eqnarray}}
\newcommand{\eeq}{\end{eqnarray}}
\newcommand{\Dslash}{\mbox{$D$\kern-0.65em \hbox{/}\hspace*{0.25em}}}
\newcommand{\Dslashup}{\not \kern-0.03em D}

\title{Improved Multiboson Algorithm}

\author{
C. Alexandrou\address{University of Cyprus, CY-1678 Nicosia, Cyprus and PSI,
 CH-5232 Villigen, Switzerland},
Ph. de Forcrand\address{ETH, CH-8092 Z\"urich, Switzerland}, 
 M. D'Elia\address{University of Pisa and INFN, I-56127 Pisa, Italy} 
and H. Panagopoulos\address{University of Cyprus, CY-1678 Nicosia, Cyprus}}
       
\begin{document}

\begin{abstract}
We compare the performance of an UV-filtered Multiboson algorithm, 
including a global quasi-heatbath update, to the standard Hybrid
Monte Carlo algorithm in full QCD with two flavours of Wilson fermions.
\end{abstract}
\maketitle

\vspace*{-1.5cm}

\section{Introduction}
 
In this study we consider QCD with 
two degenerate flavours of dynamical Wilson fermions. 
The  partition function is  given by

\be
{\mathcal Z} = \int \prod_{x,\mu} dU_{x,\mu}~e^{-S_g[U]} ~ 
 \det(\Dslash[U] + m)^2 \quad.
\ee
The fermion determinant which results after integration of the fermions
leads to non-local interactions among the
gauge links $U_{x,\mu}$ making the  cost for updating all links
grow naively like ${\mathcal O}(V^2)$, where $V$ is the lattice 
volume.

In the Hybrid Monte Carlo (HMC) algorithm, 
which is the standard method of numerical
simulations with dynamical fermions, one uses  an auxiliary 
bosonic pseudo-fermion field $\phi$ to write the determinant as
\be
|\det (\Dslash + m)|^2 = \int [d\phi^\dagger] [d\phi]~
e^{- |(\Dslashup + m)^{-1} \phi|^2} 
\label{hmceq}
\ee
and then  uses Molecular Dynamics to  globally update the $U$.

An alternative approach, which allows the use of local algorithms,
is the  Multiboson (MB) method, originally proposed by L\"uscher 
\cite{MB0}. 
A polynomial  $P_n(x) = c_n\prod_{k=1}^n (x-z_k)$ 
 is constructed  which approximates $1/x$ over the whole spectrum of  
$(\Dslash + m)$. Then  $P_n(\Dslash + m) \approx (\Dslash + m)^{-1}$
and one can write
\beq
\lefteqn{
|\det (\Dslash + m)|^2 \approx |{\det}^{-1} P_n(\Dslash + m)|^2 } \nonumber \\
&\hspace*{-0.5cm}= c_n^{-2V} \int \prod_k [d\phi_k^\dagger] [d\phi_k]~
e^{- \sum_k |(\Dslashup + m - z_k) \phi_k|^2} \quad.
\eeq
In this way the original QCD partition function is written
approximately
in terms
of a local action
so that standard powerful local algorithms (heatbath, overrelaxation)
can be used. The systematic error deriving from this approximation
can be corrected either during the simulation, with a global
accept-reject step, or by a reweighting procedure on physical observables.

We compare here the efficiency
of an improved version of the MB algorithm with 
a ``state of the art'' HMC algorithm, namely the one used by
the SESAM collaboration \cite{SESAM}. The observables used for this
comparison are the plaquette, representative of the smallest
(UV) scale, and
the  topological charge, representative of the 
largest (IR) scale features of the gauge field.

\vspace*{-0.5cm}

\section{Algorithm}

We use the exact, non-hermitian version of the MB algorithm \cite{MB1} 
with a noisy
Metropolis test to correct for the polynomial approximation
to the fermionic determinant.
There exist 
two different, complementary strategies for improving the efficiency of the MB method:\\
1.{\it UV-filtering:} 
The first is the 
choice of an optimal polynomial $P_n$ which keeps $n$ as low as
possible and still achieves a  
reasonable Metropolis acceptance.
 We  follow the procedure of \cite{PhUV} inspired by 
the loop expansion of the fermion determinant:
$ {\rm det}({\bf 1} - \kappa M) = 
 e^{- \sum_j a_j {\rm Tr} M^j} \times
{\rm det}\left( ({\bf 1} - \kappa M) 
~~ e^{+ \sum_j a_j M^j} \right) $.
We group  the term $ e^{- \sum_j a_j {\rm Tr} M^j}$ 
with the gauge action (for $j\le 4
$ this amounts to a mere shift in $\beta$)
and find a polynomial approximation to the inverse of 
$ ({\bf 1} - \kappa M) ~~e^{+ \sum_j a_j M^j}$.
The parameters have to be chosen so that $\det W \simeq 1$, where 
$ W \equiv \prod_k^n ({\bf 1} - \kappa M - z_k {\bf 1}) \cdot
({\bf 1} - \kappa M) \cdot e^{\sum_{j=0}^{m-1} a_j M^j} $. A sufficient condition for this is that $W \eta \simeq \eta$ for Gaussian vectors $\eta$.
Therefore one takes an equilibrium gauge configuration, fixes the coefficients $a_j$ to some
initial value, draws one or more Gaussian vectors $\eta$ and
finds, by quadratic minimization, the roots $\{z_k\}$ which minimize the quantity 
$e = \parallel W\eta - \eta \parallel^2 $.
The derivatives $\partial e/\partial a_j$ are also obtained, which allows
the optimization of both the coefficients $\{a_j\}$ and the zeros $\{z_k\}$,
following ref.~\cite{PhUV}.\\
2.{\it Quasi - heatbath:} 
The second is the choice of the update algorithm for gauge links and boson fields.
The coupled dynamics of the system
(gauge links + boson fields) is highly non trivial,
so that the optimal mixture of update algorithms
is essentially determined by numerical experiment \cite{Beat}.
For light quark masses one needs a good 
algorithm to speed up the IR bosonic  modes.
To implement a global heatbath on the bosonic fields one must assign
$\phi_k^{\rm NEW} = (D - z_k)^{-1} \eta $ where $D=1-\kappa M$ and 
$\eta$ is a Gaussian random vector. This procedure is  expensive especially for
small quark masses.
Instead we use a quasi-heatbath consisting of an approximate 
inversion plus a Metropolis accept-reject step as proposed in \cite{PhQH}.
i.e. we solve approximately $(D - z_k) x =  (D - z_k) \phi_k^{\rm OLD} +
\eta$ and set $\phi_k^{\rm NEW} = x - \phi_k^{\rm OLD}$.
We then 
accept with probability
\be
min \left( 1,{\rm exp} (2 Re(r^\dagger.(D - z_k) 
(\phi_k^{\rm NEW} - \phi_k^{\rm OLD}) \right)
\ee
with $r$ the residual.
We combine this global update with local overrelaxation of gauge and
bosonic fields.

\vspace*{-0.3cm}

\section{Results}

\vspace*{-0.3cm}

Three representative systems, denoted (A),(B) and (C), 
in Table I,
have been studied: medium-heavy quarks in a small lattice, light quarks in a
small lattice, and light quarks in a large lattice. 
The HMC simulations used for comparisons of these systems incorporate
state-of-the-art improvements (but without the recently implemented
SSOR preconditioning).

\begin{table}[!h]
\vspace*{-1cm}
\begin{center}
\begin{tabular}{|c|c|c|c|}
\hline
Simulation & A & B & C \\
\hline
Volume & $8^4$ & $8^4$ & $16^3 \times 24$ \\
$\beta,\kappa$ & $5.3, 0.158$ & $5.3, 0.165$ & $5.6, 0.156$ \\
$n_{bosons}$ & 7 & 16 & 24 \\
$a_2$ & 1.389 & 6.066 & 4.077 \\
$a_4$ & 4.411 & 4.423 & 8.789 \\
$\Delta\beta=192 \kappa^4 a_4$ & 0.527 & 0.629 & 0.999 \\
$\tau_{int}(\Box)(MB)$ & $\sim 3500$ & $\sim 64000$ & $\sim 27500$ \\
$\tau_{int}(\Box)(HMC)$ & $\sim 14000$ & $\sim 72000$ & $\sim 85000$ \\
\hline
\end{tabular}
\end{center}
\caption{Summary of the parameters of our 3 simulations. 
Integrated autocorrelation times are measured in units of applications of the
Wilson Dirac matrix to a vector.}
\vspace*{-1cm}
\end{table}

\begin{figure}[!h]
\vspace*{-1.cm}
\begin{center}
\epsfxsize=6.cm 
\epsfysize=5cm 
\epsfclipoff
\epsffile{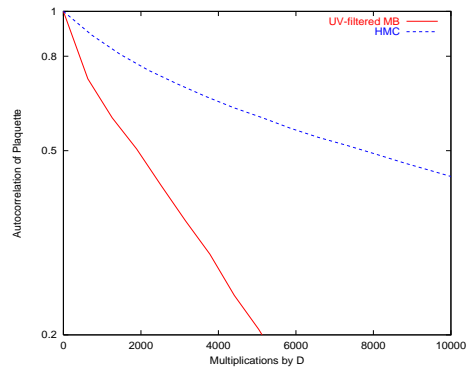}
\vspace*{-1.1cm}
\caption{ Autocorrelation of the plaquette, with
the UV-filtered MB algorithm (solid line) and the HMC algorithm
(dashed line)
at $\beta = 5.3$ and $\kappa = 0.158$ (simulation A).
}
\end{center}
\vspace*{-1.5cm}
\end{figure}

\noindent
{\it Medium - heavy quarks on a small lattice (Simulation A):}
Only seven bosonic fields are required, with only one zero $z_k$ devoted to
the UV part of the Dirac spectrum.
As demonstrated in Fig.~1
the plaquette decorrelates about 4 times
  faster than with HMC.

\noindent
{\it Light quarks on a small lattice (Simulation B):}
To see if there is a quark mass below which the MB loses its 
advantage we consider light quarks.
Several of the IR zeros 
are now closer to the Dirac
spectrum boundary.
The quasi-global heatbath provides an
  improvement of at least a factor of~5. From the plaquette autocorrelation 
 the HMC and MB algorithms appear equally efficient.

\begin{figure}[!t]
\vspace*{-0.5cm}
\begin{center}
\epsfxsize=6.cm 
\epsfysize=6.cm 
\epsfclipoff
\epsffile{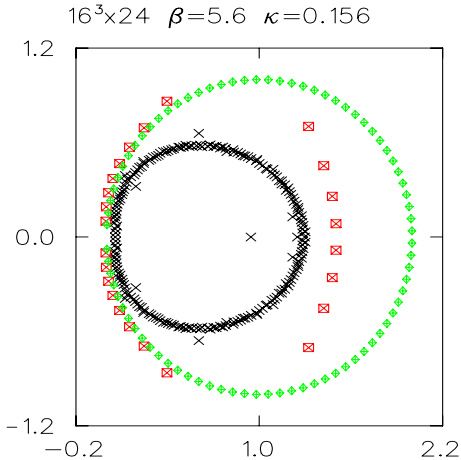}
\vspace*{-0.5cm}
\caption{Zeros of the UV-filtered polynomial ($n=24$, squares), of the
non-UV-filtered polynomial ($n=80$, diamonds), and estimated boundary
of the Dirac spectrum (x), (simulation C).}
\end{center}
\vspace*{-1.3cm}
\end{figure} 

\begin{figure}[!h]
\vspace*{-1.0cm}
\begin{center}
\epsfxsize=6.cm 
\epsfysize=5.2cm 
\epsfclipoff
\epsffile{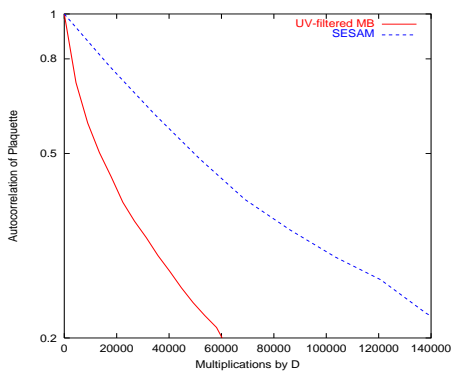}
\vspace*{-0.8cm}
\caption{Autocorrelation of the plaquette, with
the UV-filtered MB algorithm (solid line) and the SESAM HMC algorithm
(dashed line)
at $\beta = 5.6$ and $\kappa = 0.156$ (simulation C).
}
\end{center}
\vspace*{-1.5cm}
\end{figure}

\noindent
{\it Light quarks on a large lattice (Simulation C):}
The number of  boson fields needed is 24, as  compared with 80
which would have been needed with no UV-filtering
for a comparable acceptance. 
Fig.~2 shows that  16 of the 
roots of the UV-filtered polynomial are devoted to the IR modes and 8
to the UV modes of the Dirac operator.

\begin{figure}[!t]
\vspace*{-.5cm}
\begin{center}
\epsfxsize=6.cm 
\epsfysize=6.cm 
\epsfclipoff
\epsffile{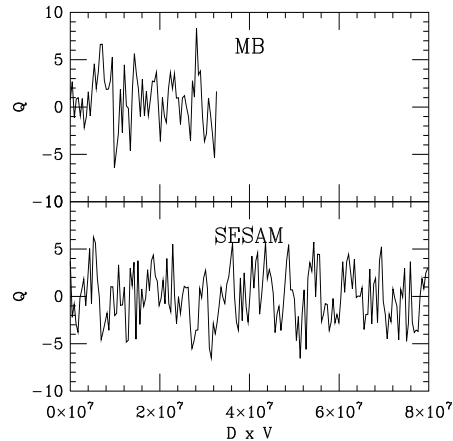}
\vspace{-1.0cm}
\caption{Comparison of topological charge histories obtained
with HMC and UV-filtered MB algorithm at $\beta = 5.6$ and $\kappa = 0.156$.
The common scale has been set in terms of equivalent $D\times v$ 
multiplications.}
\end{center}
\vspace*{-1.5cm}
\end{figure} 


The autocorrelation of the plaquette is compared in
 Fig.~3 with that obtained by the SESAM collaboration using HMC 
\cite{SESAM}.
Our MB approach is more efficient by a factor $\sim 3$.
In Fig.~4 we compare the topological charge histories obtained from
our simulation and from a sample of SESAM configurations~\cite{HMC2}.
In both cases the same cooling method was used. Neither simulation
is long enough to extract a reliable autocorrelation time for this observable.
Attempts at doing so yield roughly equivalent results for both algorithms,
as the figure already indicates. 
Therefore, even for this global observable,
our MB method seems not worse than HMC.

\vspace*{-0.3cm}

\section{Conclusions}

\vspace*{-0.3cm}

The numerical evidence presented in this study shows that
 the non-hermitian MB algorithm with  UV-filtering and 
a global  quasi-heatbath of the boson fields 
is a superior alternative to the traditional HMC:  
it decorrelates the plaquette more efficiently, and the topological
charge equivalently well.

\end{document}